\documentclass[11pt]{article}

\usepackage{varioref}
\usepackage{amsmath,amsfonts,amsthm,mathrsfs}

\usepackage[normalem]{ulem}
\usepackage{multirow,color,graphics}
\usepackage{amsmath}
\usepackage{amsfonts}
\usepackage{amssymb}
\usepackage{jheppub}
\usepackage{enumerate}
\usepackage{fancyvrb}
\usepackage{verbatim}
\usepackage{wrapfig}
\usepackage{appendix}
\usepackage{amstext}
\usepackage{amssymb}
\usepackage{graphicx}
\usepackage{color}
\usepackage{varioref}
\usepackage{multirow,graphics}
\usepackage{epstopdf}
\newcommand{\nn}{\nonumber}

\numberwithin{equation}{section}

\def\[{\left[}
\def\]{\right]}
\def\({\left(}
\def\){\right)}
\def\d{\partial}

    \newcommand{\beq}{\begin{equation}}
    \newcommand{\eeq}{\end{equation}}
    \newcommand\beqa{\begin{eqnarray}}
    \newcommand\eeqa{\end{eqnarray}}
\newcommand\bea{\begin{array}}
\newcommand\eea{\end{array}}

\newcommand{\bQ}{{\bf Q}}
\newcommand{\bq}{{\bf q}}
\newcommand{\bP}{{\bf P}}

\newcommand{\la}[1]{\label{#1}}
\newcommand{\eq}[1]{(\ref{#1})}

    \def\bQ{{\bf Q}}
    \def\bP{{\bf P}}

\makeatletter
     \@ifundefined{usebibtex}{} {}
\makeatother

    \def\bQ{{\bf Q}}

        \def\bP{{\bf P}}

\title{Quark--anti-quark potential in {\cal N=4} SYM}

\author[]{~~Nikolay Gromov${}^{\,q,\bar q}$}
\author[]{~~Fedor Levkovich-Maslyuk${}^{\,q}\ ^*$}

\affiliation[]{${}^q$Mathematics Department, King's College London,
The Strand, London WC2R 2LS, UK.}
\affiliation[]{${}^{\bar q}$St.Petersburg INP, Gatchina, 188 300, St.Petersburg,
  Russia}
\affiliation[]{${}^{*}$\!Nordita, KTH Royal Institute of Technology and Stockholm University,
Roslagstullsbacken 23, SE-106 91 Stockholm, Sweden}

\preprint{NORDITA-2016-134}

\emailAdd{nikgromov$\bullet$gmail.com}
\emailAdd{fedor.levkovich$\bullet$gmail.com}

\abstract{
We
construct
a closed system of equations describing
the quark--anti-quark potential at any coupling in planar
${\cal N}=4$
supersymmetric Yang-Mills theory.
It is based on the Quantum Spectral Curve method
supplemented
with a novel type of asymptotics.
We
present a high precision numerical solution reproducing the classical and
one-loop string predictions very accurately.
We also analytically compute the first $7$ nontrivial orders of the weak coupling expansion.

Moreover, we study analytically
the generalized quark--anti-quark potential in the limit
of large imaginary twist to all orders in perturbation theory.
We
demonstrate how
the QSC reduces in this case to a one-dimensional Schrodinger equation. In the process we establish
a link between the Q-functions and the solution of the
Bethe-Salpeter equation.
}

\begin{document}

\maketitle

\newpage

\section{Introduction}
{
\renewcommand{\thefootnote}{}
\footnote{*Alternative email for F. L.-M.: fedor.levkovich.maslyuk$\bullet$su.se}
}
\setcounter{footnote}{0}
During the recent years ${\cal N}=4$ SYM in 4d has been studied very intensively.
One of the reasons it attracted so much attention is  the existence of a well established AdS/CFT dual. At the same time it is also a beautiful theory by itself
which has numerous links with the more realistic QCD.

One of the first predictions of AdS/CFT was the strong coupling
limit of the potential between two heavy charged particles or ``quarks"
which is represented by a pair of anti-parallel Wilson lines separated by distance $r$
\cite{Erickson:1999qv,Erickson:2000af}.
The potential is inversely proportional to the separation $r$
due to conformal symmetry of the theory,
with the strength of the interaction depending on the gauge coupling $g_{YM}$.
In the planar limit $N_c\to\infty$ the potential is
a highly non-trivial function
of the 't Hooft coupling $\lambda=g^2_{YM}N_c$,
\beq
V(\lambda,r)=-\frac{\Omega(\lambda)}{r}\;.
\eeq
Currently the function $\Omega(\lambda)$ is known at $3$ loops at weak coupling
\cite{Erickson:1999qv,Pineda:2007kz,Drukker:2011za,Correa:2012nk,Bykov:2012sc,Stahlhofen:2012zx,Prausa:2013qva}
and at one loop at strong coupling \cite{Maldacena:1998im,Rey:1998ik,Forini:2010ek,Chu:2009qt}.
In fact even at low orders the weak coupling expansion is rather involved and requires using a nontrivial low-energy effective theory.
One can further generalize this observable by introducing an extra parameter $\theta$,
which may be associated with relative flavors of the particles.
The particle flavor enters through the unit vector $\vec n$
in the expression for the Maldacena-Wilson line
\beq
{\;\rm Pexp}\;\[\int(i A_\mu \dot x^\mu+\vec \Phi\cdot \vec n\;|\dot x|)\]\;.
\eeq
The parameter $\theta$ is the angle between these vectors $\vec n$
for the two antiparallel lines.
The  expectation value of the pair of the Maldacena-Wilson lines
is related to the potential as
\beq
\langle
W
\rangle\simeq e^{\frac{T\Omega(\lambda,\theta)}{r}}
\eeq
where $T\gg r$ is the extent of the lines.

In this paper we study this important observable $\Omega(\lambda,\theta)$
intensively
using
the integrability-based Quantum Spectral Curve method introduced for local operators in \cite{Gromov:2014caa,Gromov:2013pga} and generalized for a subclass of Wilson lines
in
\cite{Gromov:2015dfa}.
We show how the results of \cite{Gromov:2015dfa}  can be used to get a closed system of equations
describing $\Omega(\lambda,\theta)$ exactly in the whole range of the parameters $\lambda$ and $\theta$. We find the analytic weak coupling expansion
up to $7$ loops and also build a numerically-exact
function interpolating from weak to strong coupling regime.
Finally, we study analytically the limit $\theta\to i\infty$
(with $\lambda e^{-i\theta}$ fixed)
to all orders in the `t Hooft coupling\footnote{A similar limit in the $\gamma$-deformed SYM was recently considered in \cite{Gurdogan:2015csr}.}. We demonstrate how the
Schr\"odinger equation arising from resummation of the ladder diagrams in this limit
appears from the Quantum Spectral Curve.

\section{Quantum Spectral Curve for the quark--anti-quark potential}
The configuration of two anti-parallel Wilson lines is closely related to a configuration where
two straight lines meet at a cusp where they form an angle $\phi$ \cite{Drukker:1999zq}.
Indeed, the two setups are linked by the plane to cylinder transformation where the cusp point is mapped to infinity.
In this picture the distance between the lines is given by $r=\phi-\pi$. When $\phi$ tends to $\pi$
the curvature of the cylinder becomes irrelevant and one recovers the flat space quark--anti-quark potential.

In \cite{Drukker:2011za,Correa:2012hh,Drukker:2012de} it was shown
that the anomalous dimension of the cusped Maldacena-Wilson line
admits an integrability-based description in terms of an infinite system of
integral equations (known as Thermodynamic Bethe Ansatz equations).
This anomalous dimension
 depends on $3$ parameters: $\theta,\phi$ and the coupling $g=\frac{\sqrt\lambda}{4\pi}$.
Subsequently a much simpler description in terms of the Quantum Spectral Curve (QSC) was found \cite{Gromov:2015dfa} which we use here.

In this section we review the QSC construction
for this observable
and introduce important notation
relevant for the rest of the paper.

\subsection{QSC for a Wilson Line with a cusp}
The Quantum Spectral Curve \cite{Gromov:2014caa,Gromov:2013pga} is a set of finite difference equations
on Q-functions supplemented with very precise analytical properties.
There are $4+4$ basic Q-functions denoted as $\bP_a,\;a=1,\dots,4$
and $\bQ_i,\;i=1,\dots,4$. They can be roughly associated with
$S^5$ and $AdS_5$ degrees of freedom correspondingly.
For different observables with different quantum numbers
one should adjust accordingly the asymptotics and sometimes
modify the analytical properties. In the rest of the section
we review the QSC construction proposed in \cite{Gromov:2015dfa} for the anomalous dimension $\Delta$ of the cusped Wilson line which is a function
of two angles $\phi,\;\theta$ and the coupling $g$.

The $S^5$ part of the QSC can be efficiently parameterized in terms
of two functions ${\bf f}$ and ${\bf g}$
of the spectral parameter $u$ analytic everywhere except
the cut $[-2g,2g]$ as follows \cite{Gromov:2015dfa}:
\beqa
\label{cuspas}
\bP_1(u)&=&+\epsilon\;u^{1/2}\; e^{+\theta u}\; {\bf f}(+u)\ ,\\
\nn
\bP_2(u)&=&-\epsilon\;u^{1/2}\; e^{-\theta u}\; {\bf f}(-u)\ ,\\
\nn
\bP_3(u)&=&+\epsilon\;u^{1/2}\; e^{+\theta u}\; {\bf g}(+u)\ ,\\
\nn
\bP_4(u)&=&+\epsilon\;u^{1/2}\; e^{-\theta u}\; {\bf g}(-u)\ .
\eeqa
We have to impose ${\bf f}\simeq 1/u$ and ${\bf g}\simeq u$ for large $u$.
For this normalization the prefactor $\epsilon$
is fixed to be
\beq
\epsilon=\sqrt{\frac i2}\;
\frac{\cos \theta -\cos \phi}{\sin \theta}
\;.
\eeq
It is convenient to resolve the cut $[-2g,2g]$ by introducing the variable $x$
\beq
x(u)=\frac{u+\sqrt{u-2g}\sqrt{u+2g}}{2g}\;\;,\;\;u=(x+1/x)g\;.
\eeq
This function maps the complex plane into the exterior of the unit circle $|x|>1$.
As a regular function of $x$ at least for $|x|>1$,
the functions ${\bf f}$ and ${\bf g}$ can be written in terms of the Laurent expansion
coefficients
\beqa
\label{fgAB}
{\bf f}(u)=
\frac{1}{gx}+\sum_{n=1}^\infty \frac{g^{n-1}A_n}{x^{n+1}}
\;\;,\;\;
{\bf g}(u)=
\frac{u^2+B_0u}{gx}+\sum_{n=1}^\infty  \frac{g^{n-1}B_n}{x^{n+1}}\;.
\eeqa
The first few coefficients encode the information about the $AdS$
charges and twists, i.e. $\Delta$ and $\phi$,
via the relations \cite{Gromov:2015dfa}
\beqa
\label{epsab0}
A_1g^2-B_0&=&-\frac{ 2 \cos \theta  \cos \phi+\cos (2 \theta) -3}{2 \sin\theta (\cos \theta -\cos \phi)}\;,\\
\nn
	\Delta^2&=&
	\frac{(\cos\theta-\cos\phi)^3}{\sin\theta\sin^2\phi}\[
A_3 g^6+\frac{A_1^2 g^4 (1-\cos\theta \cos\phi)}{
\sin\theta(\cos \theta -\cos\phi)}-A_2 g^4 \cot\theta
-g^2 \left(B_0+B_1+\cot \theta\right)
\right.\\
\label{deltaas0}&&   \left.
-A_1 g^2
   \left(A_2 g^4-2 g^2+\frac{1}{\sin^2\theta}\right)
\]\;.
\eeqa
We also note that the coefficients $A_n$ and $B_n$ are real and
scale at weak coupling as ${\cal O}(g^0)$.
Their leading weak coupling behavior can be deduced from \cite{Gromov:2015dfa} and is given in Appendix~\ref{sec:WC}.

The $AdS_5$ constituents of the QSC, i.e. the $\bQ_i$, are completely fixed in terms of $\bP_a$. The procedure is the following:
first one finds the ``transition functions" $Q_{a|i}$ from the finite difference equation
\beqa\la{QQPPQ}
Q_{a|i}(u+\tfrac{i}{2})-Q_{a|i}(u-\tfrac{i}{2})=-\bP_a(u) \bP^b(u) Q_{b|i}(u+\tfrac{i}{2})
\eeqa
where $\bP^a=\chi^{ab}\bP_b$ and the
only non-zero elements of the constant matrix
$\chi^{ab}$ are $\chi^{41}=-\chi^{32}=\chi^{23}=-\chi^{14}=1$.
The finite difference equation \eq{QQPPQ}
has $4$ independent solutions labeled by the
index $i=1,\dots,4$. It is always possible to impose analyticity of $Q_{a|i}(u-\tfrac i2)$ above the real axis which we assume to be done.
Once $Q_{a|i}$ are found, $\bQ_i$ are simply given by
\beqa\la{QQP}
\bQ_i(u)=-\bP^a(u) Q_{a|i}(u+\tfrac i2)\;.
\eeqa
Equivalently these relations can be reformulated~\cite{Gromov:2015vua} in a form of a $4$th order
finite difference equation for $\bQ_i$ with the coefficients built from $\bP_a$
\beqa\la{bax5}
\bQ^{[+4]}_iD_0
&-&
\bQ^{[+2]}_i
\[
D_1-\bP_a^{[+2]}\bP^{a[+4]}D_0
\]
+
\bQ_i
\[
D_2-\bP_a\bP^{a[+2]}D_1+
\bP_a\bP^{a[+4]}D_0
\]\\
&-&
\bQ_i^{[-2]}
\[
\bar D_1+\bP_a^{[-2]}\bP^{a[-4]}\bar D_0
\]
+\bQ_i^{[-4]}\bar D_0
=0\nn
\eeqa
where $D_n, \bar D_n$ are some nice combinations of $\bP$'s given in Appendix \ref{sec:dets}.
As a $4$th order equation it has $4$ independent solutions which are precisely the
$\bQ_i$. One can show that the relations \eq{deltaas0} and \eq{epsab0} imply the following large $u$ asymptotics for $\bQ_i$
\beq
\label{cuspasQ}
	\bQ_1\sim u^{1/2+\Delta}e^{u\phi},\ \bQ_2\sim u^{1/2+\Delta}e^{-u\phi},\
	\bQ_3\sim u^{1/2-\Delta}e^{u\phi},\ \bQ_4\sim u^{1/2-\Delta}e^{-u\phi}\ .
\eeq
This is in fact how \eq{deltaas0} and \eq{epsab0} were derived.
Those $4$ distinguished asymptotics allow to choose the basis of solutions $\{\bQ_i\}$ uniquely up to a normalization.
The functions $\bQ_i$ are analytic in the upper half plane and have a cut $[-2g,2g]$
on the real axis as well as more cuts below (as can be deduced from the equation \eq{bax5}).
A non-trivial new condition, which in fact allows to close the equations
and fix the coefficients $A_n$ and $B_n$ uniquely,
concerns the behavior of $\bQ_i$ on the cut $[-2g,2g]$.
To describe it we introduce
\beq
\label{qQdef}
    {\bf q}_i=\bQ_i u^{-1/2}
\eeq
and denote by $\tilde {\bf q}_i$
the analytic continuation of ${\bf q}_i$ under the cut on the real axis.
Then according to \cite{Gromov:2015dfa}
\beqa\la{qtil}
&&\tilde {\bf q}_1(u)={\bf q}_1(-u)\\
&&\tilde {\bf q}_2(u)={\bf q}_2(-u)\\
&&\tilde {\bf q}_3(u)=a_1\sinh(2\pi u){\bf q}_2(-u)+{\bf q}_3(-u)\\
&&\tilde {\bf q}_4(u)=a_2\sinh(2\pi u){\bf q}_1(-u)+{\bf q}_4(-u)\;.
\eeqa
It was noticed in \cite{Gromov:2015dfa} that it is sufficient
to impose the first two equations in \eq{qtil} only.

In the next section we discuss what happens in the singular limit $\phi\to\pi$
and derive a closed system of equations describing directly the potential $\Omega(\lambda,\theta)$.

\subsection{QSC for the quark--anti-quark potential}
In this paper we focus on the particularly important limit $\phi\to\pi$
when the Wilson line with a cusp is related to a pair of anti-parallel lines. In this limit we expect the anomalous dimension $\Delta$
to diverge as
\beq
\Delta = -\frac{\Omega(\lambda)}{\pi-\phi}+{\mathcal{O}}((\pi-\phi)^0)
\eeq
where $\Omega(\lambda)$ is a positive quantity (for real $\theta$).
As the anomalous dimension diverges we should expect a drastic change in the large $u$
asymptotics of ${\bf Q}_i$, which for finite $\Delta$ is given by \eq{cuspasQ}.
To get some intuition about what happens we take $\phi=\pi-\epsilon$ with $\epsilon$
being small, so the asymptotics becomes
\beq
{\bf q}_1\sim e^{\pi u}\exp\[{-u\epsilon-\frac{\Omega}{\epsilon}\log u}\]\;.
\eeq
We see that the last term in the second factor explodes for $u$ fixed and the asymptotics does not make sense.
What happens is that the subleading coefficients become bigger in this limit in order to make
the result finite.
However, if we scale $u$ to infinity while sending $\epsilon\to 0$
we should be able to suppress the subleading in $1/u$ terms. The guiding principle is
to try to balance the two terms in the square brackets, which is the case for $u\sim \Omega/\epsilon^2$ (treating $\log u$ as a constant
compared to $\sqrt u$)
or in other words for $\epsilon = c \sqrt\Omega/\sqrt{u}$ this results in
\beq\la{qasm}
\log {\bf q}_1\sim +\pi u-c\sqrt{\Omega u}+{\cal O}(u^0)\;.
\eeq
The positive constant $c$ cannot be determined from this heuristic argument
and it will be shown below to be equal to $\sqrt{8}$.
Similar considerations for ${\bf q}_2, {\bf q}_3$ and  ${\bf q}_4$ lead to
\beqa
&&\log {\bf q}_2\sim -\pi u+i c\sqrt{\Omega u}+{\cal O}(u^0)\;,\\
&&\log {\bf q}_3\sim +\pi u-i c\sqrt{\Omega u}+{\cal O}(u^0)\;,\\
&&\log {\bf q}_4\sim -\pi u+ c\sqrt{\Omega u}+{\cal O}(u^0)\;.
\eeqa

To get the precise value of the coefficient $c$ and derive the asymptotics rigorously, we have to analyze the limit of $\bP_a$ when $\phi\to\pi$.
One could expect that $\bP_a$ behave smoothly in this limit
as they describe the $S^5$ part which is relatively isolated from the twist $\phi$ in $AdS_5$.
It can be also seen from \eq{epsab0} and \eq{deltaas0} that we can consistently assume the coefficients $A_n$
and $B_n$ in $\bP$'s to remain finite when $\phi\to\pi$, giving
\beqa
\epsilon&=&\sqrt\frac{i}{2}\frac{\cos \theta +1}{\sin \theta}\;\;,\;\;B_0=A_1g^2-\frac{2-\cos\theta}{\sin\theta}\;,\\
\nn\la{omegAB}
	\Omega^2&=&
\frac{g^2 \cot ^3\frac{\theta }{2}}{2}
\[
2 \sin\theta  \left(A_3 g^4 \sin\theta -A_2 g^2 \cos \theta -B_1
   \sin\theta -2 \cos \theta+2\right)\right.\\
&&\left.+2 A_1^2 g^2 \sin \theta+A_1 \left(-2 A_2 g^4 \sin ^2\theta-g^2 \cos (2 \theta
   )+g^2-2\right)
\]\;.
\eeqa
This allows to find the asymptotics of ${\bf q}_i$ using the
4th order Baxter equation \eq{bax5} in which we expand the coefficients at large $u$.
The expressions we get are lengthy, and for illustration purposes let us drop some of the terms which do not affect the leading asymptotics, leaving the following equation:
\beqa\nn
	{\bf q}(u) \left(\frac{-\frac{2 \Omega ^2}{3}-1}{u^2}+1\right)&+&\left(-\frac{2}{3 u^2}+\frac{i}{3 u}+\frac{2}{3}\right)
   {\bf q}(u+i)+\left(-\frac{1}{6 u^2}+\frac{i}{6
   u}+\frac{1}{6}\right) {\bf q}(u+2 i)
\\
\nn&+&\left(-\frac{2}{3 u^2}-\frac{i}{3
   u}+\frac{2}{3}\right) {\bf q}(u-i)+\left(-\frac{1}{6 u^2}-\frac{i}{6 u}+\frac{1}{6}\right) {\bf q}(u-2 i)=0\;.
\eeqa
While the coefficients in this equation are simple, the asymptotics of its four solutions is quite nontrivial. It turns out to have indeed the form anticipated above in \eq{qasm} as we get
\beq
	\bq_i= M_i u^{1/4}e^{\pm \pi u+\alpha_i\sqrt{u}}
	\(1+{\cal O}(1/u)\)
\eeq
where
\beq
	\alpha_1=-\sqrt{8\Omega},\ \alpha_2=+i\sqrt{8\Omega},\ \alpha_3=-i \sqrt{8\Omega},\ \alpha_4=+\sqrt{8\Omega}\;.
\eeq
Expanding the Baxter equation to higher orders in $u$ and keeping all the terms, we found the following expansion for the solution:
\beq\la{asymptotics}
	\bq_i= M_i u^{1/4}e^{\pm \pi u+\alpha_i\sqrt{u}}
	\(1+\sum_{n=1}^\infty \frac{d_n}{(\alpha_i)^n u^{n/2}}\)
\eeq
	
This rather surprising asymptotics is a key result which supplements the QSC functional equations.

A natural way to fix the normalization of $\bQ_i$ (which we will use in this paper, e.g. in Appendix \ref{sec:compcon})
is to impose that the matrix $Q_{a|i}$ preserves the constant matrix $\chi^{ab}$, i.e.
\beq
Q_{a|i}\chi^{ab}Q_{b|j}\chi^{jk}=\delta^{k}_i
\eeq
This leads to
\beq\la{MMnorm}
	i M_1M_4= M_2M_3=\sqrt 2\frac{\cos^4(\theta/2)}{\Omega^{3/2}}\;.
\eeq

We conclude that the quark--anti-quark potential is described by QSC with a novel type of asymptotics of the Q-functions
containing non-integer powers of the spectral parameter $u$ in the exponent. These asymptotics together with the general
relations from the previous section form a closed system of equations applicable at all values of the
coupling $g$ and the twist $\theta$.

Despite the anomalous dimension $\Delta$ of the cusped Wilson line being infinite at $\phi=\pi$,
we managed to reformulate the QSC equations in such a way that they only include the finite residue $\Omega(\lambda,\theta)$
and got rid of the auxiliary parameter $\phi$ completely.
In the following sections we will solve these equations both analytically at weak coupling to a high order
and numerically in a wide range of the coupling.
We will be also able to demonstrate how in a special limit the QSC reduces to the Schr\"odinger equation of \cite{Erickson:1999qv,Correa:2012nk} resumming the ladder diagrams
 to all orders in perturbation theory.

\section{Weak coupling}\la{sec4}
In this section we show how to solve the equations from the previous section
perturbatively at weak coupling. We will see that the weak coupling limit
is rather nontrivial and contains qualitatively new features compared to all other perturbative expansions
of the Quantum Spectral Curve studied previously \cite{Gromov:2013pga,Marboe:2014gma,Gromov:2015vua,Anselmetti:2015mda}.

\subsection{Different scales and structure of the expansion}\la{subsec:weak}
The weak coupling limit is more involved in the present case as
it depends on the scaling of the spectral parameter $u$.
The situation here is similar to the conventional perturbation theory
where in order to compute the quark--anti-quark potential
one has to work with an effective theory resumming soft contributions.
We also note that the limits $\phi\to\pi$ and $g\to 0$ do not commute with each other
and it is crucial to have a closed system of equations directly at $\phi=\pi$
in order to get a sensible weak coupling expansion.

Another key feature of the weak coupling calculation is that
the limits $g\to 0$ and $u\to\infty$ do not commute.
The reason for this is that $\Omega$, appearing in the asymptotics \eq{asymptotics}, goes to zero as $g^2$.
In this case one should expect the following three natural scales
\beqa
\nn&&{\rm scale\;1:\;}u\to \infty\;\;{\rm before}\;\;g\to 0\\
\nn&&{\rm scale\;2:\;}g\to 0\;\;{\rm with}\;\;v\equiv 8u\;\Omega\;\;{\rm fixed}\\
\nn&&{\rm scale\;3:\;}g\to 0\;\;{\rm then}\;u\to\infty
\eeqa
In the scale $1$ we are in the regime where
the asymptotics \eq{asymptotics} is still valid.
The scale 2 is natural to consider as in the asymptotics \eq{asymptotics} $u$ appears in this combination with $\Omega$.
In the scale $3$ we are in the usual perturbative regime of the QSC studied
intensively in \cite{Gromov:2013pga,Marboe:2014gma,Gromov:2015vua,Anselmetti:2015mda} and we should expect the usual expansion of the Q-functions
in terms of $\eta$-functions introduced in \cite{Leurent:2013mr}. These $\eta$-functions are defined as\footnote{In some
cases the sum could be divergent,
we regularize it as in \cite{Leurent:2013mr,Marboe:2014gma} so that e.g. $\eta_1(u)=i\psi(-iu)$.}
\beq
\eta_{s_1,\dots,s_k}(u)=
\sum_{n_1>n_2\dots n_k\geq 0}
\frac{1}{(u+in_1)^{s_1}\dots (u+in_k)^{s_k}}\;.
\eeq
At large $u$ however these functions can only give terms of the type $u^n\log^m u$,
which are very different from the scale 1.
The intermediate scale $2$ should match the two regimes corresponding to scales $1$ and $3$.
This regime plays an important role as it allows to identify correctly
$\bq_1$ and $\bq_2$ in the scale $3$ and distinguish them
from  $\bq_3$ and $\bq_4$, for which the
analyticity condition on the cut $[-2g,2g]$ given by \eq{qtil}
is different.

Thus, before we can use \eq{qtil} and fix the coefficients $A_n$ and $B_n$
in the expression for $\bP_a$, we have to pass through the regime with finite
$v\equiv 8\Omega u$. Fortunately, in this regime
the finite difference equation \eq{bax5} on $\bq_i$ (related to $\bQ_i$ via \eq{qQdef}) simplifies into a $4$th order differential equation
which we can solve systematically order by order in $g$.
Its solution provides a bridge between scale 1 and scale 3
by interpolating between the exponential and power-like with logs asymptotics.
We will first demonstrate this procedure at the leading order in the coupling and
then present our result to a high order in perturbation theory\footnote{At high orders to simplify intermediate expressions we used the HPL Mathematica package \cite{HPL} and the package for working with multiple zeta values provided with the paper \cite{Leurent:2013mr}. }.

To study the 2nd scale we start from the $4$th order Baxter equation \eq{bax5}
and expand it at large $u$ (notice that in the 2nd scale $u$ is large as it is $\sim1/g^2$). By doing this we obtain a finite difference equation of the
form
\beqa
{{\bf q}_i(u)}\(1+\frac{C_0}{u^2}+\dots\)&=&\frac{{\bf q}_i(u+2i)}{6}\(1+\frac{i}{u}+\frac{C_2}{u^2}+\dots\)+
\frac{2{\bf q}_i(u+i)}{3}\(1+\frac{i}{2u}+\frac{C_1}{u^2}+\dots\)\nn
\\ \nn &
+&\frac{{\bf q}_i(u-2i)}{6}\(1-\frac{i}{u}+\frac{\bar C_2}{u^2}+\dots\)+
\frac{2{\bf q}_i(u-i)}{3}\(1-\frac{i}{2u}+\frac{\bar C_1}{u^2}+\dots\)
\eeqa
where $C_n$ and the sub-leading coefficients are some explicit combinations of $A_n$ and $B_n$.
Next we use that $u=v/(8\Omega)$ where $\Omega\sim g^2$ and
introduce a smooth function $f(v)$ such that
${\bf q}(u)=e^{\pm\pi u}f(8\Omega u)$ to obtain
\beq\la{feqn}
f^{(4)}+\frac{2 f^{(3)}}{v}-\frac{f}{16
   v^2}+8 \hat g^2 \frac{f''}{v^2}+{\cal O}\left(g^4\right)=0
\eeq
 where $\hat g = g \cos\!\left(\tfrac{\theta
   }{2}\right)$.
Fortunately, we can solve this equation analytically!
At the leading order in $g$
its $4$ independent solutions are given by four different types of Bessel functions,
\beq
\sqrt{v}\;K_1(\sqrt v)\;\;,\;\;
\sqrt{v}\;Y_1(\sqrt v)\;\;,\;\;
\sqrt{v}\;I_1(\sqrt v)\;\;,\;\;\sqrt{v}\;J_1(\sqrt v)\;.
\eeq
Next we notice that the first solution
should be related to $\bq_1$ simply because
its large $v$ asymptotics matches precisely the asymptotics \eq{asymptotics} of $\bq_1$:
\beq
f_1(v)\equiv\sqrt{v}\;K_1(\sqrt v)\simeq \sqrt\frac{\pi}2 \sqrt[1/4]{8\Omega u}
\;\;e^{-\sqrt{8\Omega u}}\;.
\eeq
We note that since this is one of the decaying ``small"
solutions this identification is non-ambiguous.

At the higher orders in $g$
the equation \eq{feqn} gets corrected.
In general one would have to solve
\eq{feqn} using perturbation theory,
involving Green's function and multiple integrations.
However, we found a much simpler procedure,
which works magically up to at least $g^{10}$
order. Once can simply build an ansatz for the corrected solution
as a linear combination of $v^{(1/2-m)} \d_\nu^n K_\nu(\sqrt v)|_{\nu=1}$
for integer $m$ and $n$.
So for instance at $g^2$ order we simply get
\beq\la{f1sol}
f_1(v)=\sqrt{v}\;K_1(\sqrt v)-8\hat g^2 \sqrt{v}\;
K^{(1,0)}_1(\sqrt v)+{\cal O}({g^4})\;.
\eeq

Having an explicit form of the solution in the scale 2,
we can get information about the behavior of $\bq_1$
in the scale $3$.
For that we expand \eq{f1sol} at small $v$,
\beqa\la{f1small}
f_1(v)&=&1+\frac{1}{4} v \(\log \frac{v}{4}+2 \gamma -1\)+
\frac{1}{64} v^2 \(2 \log \frac{v}{16}+4 \gamma-5\)+O\left(v^{3}\right)\;\\
\nn&&\;\;\;\;+4 \hat g^2 \(\log \frac{v}{4}+2 \gamma \)\;\;\;\;\;\;+\hat g^2 v \(\log\frac{v}{4}+2 \gamma -2\)
+O\left(g^2v^2\right)+{\cal O}(g^4)\;.
\eeqa
We see that this expansion,
rewritten in terms of $u$ gives the {\it large} $u$
expansion of $\bq_1$ in the scale 3.
So the first line (originating from the leading order
in $g$ in \eq{f1sol})
gives the leading large $u$ term to all orders in $g$
in this scale,
the second line in \eq{f1small}
gives the subleading in large $u$ term to all orders in $g$ etc.
This information is essential for the correct identification of ${\bf q}_1$
in the scale $3$.

Now let us finally
describe the situation in the scale $3$. In this scale the
$4$th order finite difference equation cannot be much simplified
but it can be solved iteratively order by order in the coupling $g$ using
the highly universal procedure from \cite{Gromov:2015vua}. For instance at the first two orders we start by finding
$4$ independent solutions for $q=\bq \,e^{\pm\pi u}$,
\beqa
\nn q_{I}&=&1+g^2 \left(4 iu\; \eta _2 \cos ^2\frac{\theta }{2}+2\; \eta _1 \cot
   ^2\frac{\theta }{2}((u+i) \cos \theta+u-i)
	\right. \\ \nn &&\left.
	+\frac{\cot
   ^2\frac{\theta }{2}\left(2 u^3 \cos\theta+2 u^3-2
   u-i\right)}{u}\right)\;,\\
\nn q_{II}&=&u\;,\\
\nn q_{III}&=&u^2\;,\\
q_{IV}&=&4 \eta _1 u \cos ^2\frac{\theta }{2}-\frac{i}{u}\;.
\eeqa
However, to be able to use the key analyticity condition \eq{qtil}
we need to identify ${\bf q}_1$ (or ${\bf q}_2$).
That is, we have to find
a linear combination of $q_{I},\dots,q_{IV}$
which matches \eq{f1small} at large $u$.
From this condition one finds uniquely
\beq
{\bf q}_1=e^{\pi u}\(
A_{I}q_{I}+A_{II}q_{II}
+A_{III}q_{III}
+A_{IV}q_{IV}
\)+{\cal O}(g^4)
\eeq
where
\beqa
A_{I}&=&1+\hat g^2 \left(4 \log \left(2 \Omega \right)+2 \csc
   ^2\frac{\theta }{2}+i \frac{\Omega}{\hat g^2} +2 \pi i-4+8 \gamma \right)\\
A_{II}&=&0+\Omega (2 \log (2 \Omega )+i \pi +4 \gamma -2)\\
A_{III}&=&0-\hat g^2\left( 4\cot ^2\frac{\theta }{2}\right)\\
A_{IV}&=&
0-\hat g^2 \left(\csc ^2\frac{\theta }{2}
+\frac{i\Omega}{2\hat g^2}  \sec ^2\frac{\theta
   }{2}\right)\;.
\eeqa
In this way we deduce ${\bf q}_1$. This allows us to
find $\tilde {\bf q}_1(u)={\bf q}_1(-u)$ via \eq{qtil}.
On the last step of the procedure we consider the
combinations
\beq\la{regular}
{\bf q}_1(u)+\tilde{\bf q}_1(u)\;\;,\;\;
\frac{{\bf q}_1(u)-\tilde{\bf q}_1(u)}{\sqrt{u^2-4g^2}}
\eeq
in which the cut on the real axis disappears. As at weak coupling the cuts manifest themselves as poles, thus the poles at the origin which are naturally present in $\bq_1$
should cancel in these combinations \cite{Gromov:2013pga}.
This condition fixes
the coefficients $A_n$ and $B_n$ and also the value of the energy
at the given order in $g$.
So, for instance, at the $g^2$ order
we find the following
expansion at the origin
\beq
{\bf q}_1(u)\simeq \[1+\pi u+{\cal O}(u^2)\]-\Omega\[
-\frac{\sec^2\frac\theta2}{2u}+{\cal O}(u^0)
\]+{\cal O}(g^4)\;.
\eeq
Then regularity of the second combination in \eq{regular} relates
the singular term proportional to $\Omega$
with the linear coefficient $\pi u$ so that we get
\beq
\Omega=4\pi g^2\cos^2\frac\theta2+{\cal O}(g^4)\;.
\eeq
This perfectly matches the well known leading order result.

\subsection{Expansion to high order in the coupling}
The procedure described above allows to efficiently generate the quark--anti-quark potential expanded to very high orders in $g$.
We have computed the expansion up to $g^{14}$ order.
The result up to $g^{10}$ order is shown below
\beqa\la{omegapert}
\frac{\Omega}{4\pi}&=&\hat g^2+\\
\nn&&\hat g^4\[16L-8\]+\\
\nn&&\hat g^6\[128 L^2+L \left(64+\frac{64 \pi ^2 T}{3}\right)-112-\frac{8 \pi ^2}{3}+72 T \zeta_3\]+\\
\nn&&\hat g^8\[\frac{2048 L^3}{3}+\frac{1024}{3} \pi ^2 L^2 T+2048 L^2+L T \left(768 \zeta_3+\frac{2176
   \pi ^2}{3}\right)+\left(-768-\frac{640 \pi ^2}{3}\right) L\right.\\
\nn&&\left.+T^2 \left(128 \pi ^2 \zeta_3-760 \zeta_5\right)+T \left(384 \zeta_3-640 \pi ^2+\frac{32 \pi
   ^4}{9}\right)+\frac{1664 \zeta_3}{3}+\frac{1216 \pi ^2}{9}-1280\]+\\
\nn&&\hat g^{10}\[
\frac{8192 L^4}{3}+\frac{8192}{3} \pi ^2 L^3 T+\frac{57344 L^3}{3}+\frac{2048}{9} \pi ^4
   L^2 T^2+L^2 T \left(3072 \zeta_3+\frac{71680 \pi
   ^2}{3}\right)\right.\\
\nn&&+\left(20480-\frac{19456 \pi ^2}{3}\right) L^2+L T^2 \left(\frac{8704 \pi
   ^2 \zeta_3}{3}-6400 \zeta_5+\frac{2560 \pi ^4}{3}\right)\\
\nn&&+L T \left(12800 \zeta_3-\frac{46592 \pi ^2}{3}-\frac{6656 \pi ^4}{45}\right)+L \left(\frac{26624 \zeta_3}{3}-26624+\frac{38912 \pi ^2}{9}\right)\\
\nn&&+T^3 \left(\frac{1792 \pi ^4 \zeta_3}{45}-\frac{4928 \pi ^2 \zeta_5}{3}+8624 \zeta_7\right)\\
\nn&&+T^2 \left(3392 \pi ^2
   \zeta_3+1248 \zeta_3^2-4000 \zeta_5-\frac{1024 \pi ^4}{3}-\frac{16 \pi
   ^6}{45}\right)\\
\nn&&+T \left(896 \zeta_3+\frac{3392 \pi ^2 \zeta_3}{3}+1600 \zeta_5-\frac{10112 \pi ^2}{3}+\frac{1408 \pi ^4}{45}\right)\\
\nn&&\left.+6656 \zeta_3+\frac{736 \pi
   ^4}{45}+\frac{5824 \pi ^2}{27}-\frac{37888}{3}
\]\ .
\eeqa
Here we use the following notation
\beq
\hat g \equiv g\cos\frac\theta2\;\;,\;\;T
\equiv \frac{1}{\cos^2\frac\theta2}\;\;,\;\;L \equiv  \log
\sqrt{8 e^{\gamma } \pi  \hat g^2}\;.
\eeq
In Appendix \ref{loop6}
we also give the expression for the quite lengthy $\hat g^{12}$ and $\hat g^{14}$ orders. They are particularly interesting since at order $\hat g^{12}$ an irreducible multiple zeta value appears for the first time (namely, $\zeta_{6,2}$). We also give the full 7-loop result in a Mathematica notebook attached to this paper.

We notice that at the $g^{2n+2}$ order the result is a $n$th order polynomial in $L$ and $T$.
The terms with the maximal power of $L$ and the subleading in $L$
terms have a very simple structure which
can be summarized by the following formula
\beq
\label{leadlogs}
\frac{\Omega}{4\pi}=
\sum_{n=0}^\infty\hat g^{2n+2} \frac{16^n L^n}{n!}\(1+\frac{3n^2-5n}{4 L}+\pi ^2T\frac{n^2-n}{12 L}
+{\cal O}(1/L^2)
\)\;.
\eeq

Our 7-loop result computed from the QSC is in perfect agreement with direct field theory perturbative calculations.
The first three orders were known completely and were computed in \cite{Erickson:1999qv,Pineda:2007kz,Drukker:2011za,Correa:2012nk,Bykov:2012sc,Stahlhofen:2012zx,Prausa:2013qva}. In addition, our formula \eq{leadlogs} matches the all-orders prediction of \cite{Pineda:2007kz} for the coefficients of the leading logarithmic terms $\hat g^{2n}\log^{n-1}\hat g$. We also reproduced\footnote{Some of the perturbative field theory calculations discussed here were done for the special case $\theta=0$ only.}  the result of \cite{Stahlhofen:2012zx} for the subleading logaithmic term at 4th nontrivial order (i.e. $\hat g^{8}\log^{2}\hat g$).

In the next section we will show that the terms which do not contain
$T$ can be captured by a much simpler set of equations.

\section{Ladders limit of the quark--anti-quark potential}
A remarkable special limit, revealing rich structures,  is the ``double scaling" limit when the twist $t=e^{i\theta/2}$
scales to zero as $g$. In this limit the effective coupling
$\hat g = \frac{g}{2t}(1+t^2)$
and $\Omega(\hat g)$
remain finite. It is expected that in this special case our system of equations
can be solved exactly to all orders in $\hat g$ or at least simplified considerably.
From the gauge theory side, only the ladder diagrams contribute
in that limit. Their resummation is achieved by Bethe-Salpeter techniques  which
results in a Schr\"odinger equation
 \cite{Erickson:1999qv,Correa:2012nk}
\beq
F''(z)+F(z)\(
\frac{4\hat g^2}{z^2+1}
-\frac{\Omega^2}{4}
\)=0\;,
\eeq
whose ground state energy gives the quark--anti-quark potential $\Omega(\hat g)$.
Its expansion in small $\hat g$ should capture all terms in \eq{omegapert} without $T$
to all orders in $\hat g$, as $T\to 0$ in this limit.
Below we will demonstrate how this Schr\"odinger equation is encoded into the QSC.

\subsection{Double scaling limit of the QSC}
The main simplification in this limit occurs
because $g\to 0$ and thus each of the cuts $[-2g,2g]$ collapses into a point.
In particular this implies that ${\bf f}(u)$ and ${\bf g}(u)$ from \eq{cuspas}, as analytic functions
everywhere except the cut, reduce to simple rational functions.
Nevertheless, the result is a nontrivial function of the coupling $\hat g$
which resums the usual perturbative expansion.
In this sense this setup reminds the BFKL limit of the QSC studied in \cite{Alfimov:2014bwa,Gromov:2015vua}.
Special care should be taken with the exponents $e^{\pm \theta u}$ in $\bP_a$
which give extra factors of $t$ or $1/t$ each time we shift the argument $u$ by $\pm i/2$.
For this reason we have to keep terms up to order $t^4$ in $\bP_a$.
Assuming all the coefficients $A_n,B_n\sim 1$ (which we initially deduced from
the weak coupling solution described in Sec. \ref{sec4},
and confirmed by self-consistency) we get

\beqa\la{fgscale}\nn
{{\bf f}(u) }&=&
\frac{1}{u}+\frac{4 \hat{g}^2 t^2 \left(A_1 u+1\right)}{u^3}+\frac{8 \hat{g}^2 t^4
   \left(2 \hat{g}^2 \left(A_2 u^2+2 A_1 u+2\right)-u^2 \left(A_1
   u+1\right)\right)}{u^5}
	+{\cal O}(t^6)
	\\
{\bf g}(u)&=&\nn
u-i+t^2 \left[\frac{4 \hat{g}^2 \left(A_1 u^2+B_1+u-i\right)}{u^2}+4 i\right]
+t^4
   \left[-\frac{8 \hat{g}^2 \left(A_1 u^2+B_1+u-3 i\right)}{u^2}
	\right. \\
	&+&\left.
	\frac{16 \hat{g}^4
   \left(A_1 u^2+\left(B_2+2\right) u+2 \left(B_1-i\right)\right)}{u^4}-2
   i\right]
	+{\cal O}(t^6)
\eeqa
We can also exclude $B_1$ using the expression for $\Omega$ \eq{omegAB},
\beq
B_1=2 i+t^2 \left(4 A_1 \hat{g}^2+4 i A_2 \hat{g}^2-\frac{i
   \Omega ^2}{\hat{g}^2}-4 i\right)+{\cal O}\left(t^4\right)\;.
\eeq
Next we plug the expressions \eq{fgscale} into \eq{bax5} and expand to the leading order in $t$.
We notice that the dependence on all remaining $A_n$ and $B_n$ disappears and we simply get
\beqa\la{bax4o}
&& \nn
\left(\frac{16
   \hat g^4}{u^3}+\frac{16
   \hat g^2}{u}-\frac{4\Omega^2}{u}+6
   u\right)q(u)
	\\
	&+&
(u+i) q(u+2 i)
-\left(\frac{4 \hat g^2 (2 u+i)}{u (u+i)}+ 4
   u+2i\right) q(u+i)\\ \nn
&+&
(u-i) q(u-2 i)
-\left(\frac{4 \hat g^2 (2 u-i)}{u (u-i)}+ 4
   u-2i\right) q(u-i)=0
\eeqa

where $q(u)=\bQ(u)e^{\pm\pi u}/\sqrt u$.
A great simplification comes from the fact that this equation can be factorized into two second order equations!
This allows to replace \eq{bax4o} by a pair of second order equations
\beqa\la{bax2}
\boxed{-2q(u)\(
2\hat g^2-\Omega u
+u^2
\)+u^2 q(u-i)
+u^2 q(u+i)=0}
\eeqa
and the second one related by $\Omega\to -\Omega$.
By analyzing the large $u$ asymptotics it is easy to see
that the two solutions of \eq{bax2} correspond to $\bQ_1$
and $\bQ_4$. To fix the conventions and normalizations we define
\beq\la{q1q4}
q_1\simeq
\sqrt{\pi/2} \sqrt[1/4]{8\Omega u}
\;e^{-\sqrt{8\Omega u}}
\;\;,\;\;
q_4\simeq
\frac{1}{16i\pi\Omega^2 t^4}\sqrt{\pi/2} \sqrt[1/4]{8\Omega u}
\;e^{+\sqrt{8\Omega u}}
\;\;,\;\;
q_4(0)=0
\eeq
where
\beq
q_1 = e^{-\pi u}\bQ_1/\sqrt u\;\;,\;\;q_4 = e^{+\pi u}\bQ_4/\sqrt u\;.
\eeq
The relative coefficient in \eq{q1q4} is chosen in agreement with the canonical normalization \eq{MMnorm}.
We also choose $q_1$ and $q_4$ to be regular in the upper half plane as usual.
We see that \eq{bax2} is invariant under complex conjugation,
which implies that $\bar q_1$ and $\bar q_4$ are some linear combinations
of $q_1$ and $q_4$
with $i$-periodic coefficients
\beqa\la{barq}
\bar q_1&=&
\Omega_{1}^{\;1} q_1+
e^{-2\pi u}\Omega_{1}^{\;4} q_4\\
\bar q_4
&=&
e^{+2\pi u}\Omega_{4}^{\;1}q_1
+\Omega_{4}^{\;4}q_4\;.
\eeqa
Here $\Omega_i^{\;j}$ are some  $i$-periodic functions for which notation is introduced in accordance with the general
consideration from Appendix \ref{sec:compcon}.
Knowing the analytical properties of $q_1$ and $q_4$, which follow from the equation \eq{bax2},
we can constrain the possible form of $\Omega_i^{\;j}$.
From the equation \eq{bax2}
we can see that $q_1$ should have double poles at $u=-2i n$ for $n=1,2,\dots$ due to the $u^2$ factors in the equation.
Similarly $q_4$ has simple poles at the same points due to the additional condition $q_4(0)=0$ which softens the singularity.
Furthermore, the complex conjugate functions $\bar q_1$ and $\bar q_4$
should have the same poles as $q_1$ and $q_4$ but in the
upper half-plane instead of the lower half-plane.
The poles of $\bar q_1$ in the upper half plane can only originate
from $\Omega's$ in the r.h.s. of \eq{barq}. This implies that
$\Omega_{1}^{\;1}$ and $\Omega_{1}^{\;4}$ can have at most $2$nd order poles, similarly,
 $\Omega_{4}^{\;1}$ and $\Omega_{4}^{\;4}$ can only have simple poles.
Next, if we expand \eq{barq} near $u=0$ in order to cancel poles in the r.h.s.
we must assume that $\Omega_{1}^{\;1}$ has simple pole only as $\bar q_4(0)=0$.
Similarly $\Omega_{4}^{\;1}$ should be regular.
Finally, since for large $u$ the asymptotics of $q_1$ does not contain periodic exponents
due to the definition \eq{q1q4}
we can write the following ansatz for $\Omega's$ in terms of a few constants $a_i$:
\beqa
\Omega_{1}^{\;1}=\frac{a_1+a_2 e^{2\pi u}}{e^{2\pi u}-1}\;\;,\;\;
\Omega_{1}^{\;4}=\frac{a_3 e^{2\pi u}+a_0}{(e^{2\pi u}-1)^2}
\;\;,\;\;
\Omega_{4}^{\;1}=a_4 e^{-2\pi u}
\;\;,\;\;
\Omega_{4}^{\;4}=\frac{a_5+a_6 e^{2\pi u}}{e^{2\pi u}-1}\;.
\eeqa
We also note that $a_0=0$ since $\Omega_{1}^{\;4}$ should be even as explained in \eq{dom14}.
 By comparing the large $u$ asymptotics in the first equation of
\eq{barq} at $u\to-\infty$ we can fix $a_3$ and get $\Omega_{1}^{\;4}$
\beq\la{O14}
a_3={16\pi t^4\Omega^2}\;\;,\;\;\Omega_{1}^{\;4}=\frac{4\pi t^4\Omega^2}{\sinh^2(\pi u)}\;.
\eeq
This allows to close the equations.
Indeed, by rewriting
\beq
\Omega_{1}^{\;4}=\[\frac{\tilde\Omega_{1}^{\;4}+\Omega_{1}^{\;4}}{2}\]-\[\frac{\tilde\Omega_{1}^{\;4}-\Omega_{1}^{\;4}}{2\sqrt{u^2-4g^2}}\]\sqrt{u^2-4g^2}
\eeq
so that the expressions in the square brackets are regular at the origin to all orders in $g$ we see
that the poles present in \eq{O14} can only originate from the last term.
At the same time the last term can be written in terms of $q$ and $\bar q$ using \eq{dom14}:
\beq
\[\frac{\tilde\Omega_{1}^{\;4}-\Omega_{1}^{\;4}}{2\sqrt{u^2-4g^2}}\]=
-\frac{u\bar{q}_1(-u)q_1(-u)e^{-2\pi u}
-
u\bar{q}_1(u)q_1(u)e^{+2\pi u}}{2u}+{\cal O}(g^2)=b u+{\cal O}(u^3)+{\cal O}(g^2)
\eeq
which results in the following pattern of the leading singularities in $\Omega_1^{\;4}$
\beq
\Omega_1^{\;4}= \frac{2b g^4}{u^2}+\frac{4 b g^6}{u^4}+\dots +{\text{less singular terms}}
\eeq
thus we can relate $b$ to $\Omega(\hat g)$ as
\beq
b=\frac{\Omega^2(\hat g)}{8\pi \hat g^4}
\eeq
or
\beq\la{quantization}
\boxed{
\frac{\Omega^2(\hat g)}{8\pi \hat g^4}=\lim_{u\to 0}\frac{\bar{q}_1(u)q_1(u)e^{+2\pi u}-\bar{q}_1(0)q_1(0)}{u}\;.
}
\eeq
This condition together with the finite difference equation
\eq{bax2} allows to determine $\Omega(\hat g)$.
Namely,
we have to find such value of the parameter $\Omega$ in the finite difference equation \eq{bax2}
for which its solution $q_1$ with the asymptotic \eq{q1q4},
expanded at the origin, satisfies the condition \eq{quantization}.
This type of  problem can be easily solved numerically or perturbatively
in $\hat g$.

To solve the system perturbatively we repeat basically the same steps as in the previous
section, with an additional simplification that we do not have to tune
any parameters in $\bP_a$ except $\Omega(\hat g)$, and that
we only have to deal with the second order equation
instead of the $4$th order equation.
This procedure, explained in detail in Sec.~\ref{subsec:weak}, leads to the following result
\beqa
&&\frac{\Omega(\theta= i \infty)}{4\pi}=\hat g^2+\\
\nn&&\hat g^4\[16L-8\]+\\
\nn&&\hat g^6\[128 L^2+64L-112-\frac{8 \pi ^2}{3}\]+\\
\nn&&\hat g^8\[\frac{2048 L^3}{3}+2048 L^2
-\left(768+\frac{640}{3} \pi ^2\right)L-1280+\frac{1216}{9}\pi ^2+\frac{1664}{3} \zeta_3\]+\\
\nn&&\hat g^{10}\[
\frac{8192 L^4}{3}+\frac{57344 L^3}{3}+\left(20480-\frac{19456 \pi ^2}{3}\right) L^2-
   \left(26624-\frac{38912 \pi ^2}{9}-\frac{26624 \zeta_3}{3}\right)L\right.\\
\nn&&\left.
\;\;\;\;\;\;-\frac{37888}{3}
+\frac{5824 \pi ^2}{27}
+6656 \zeta_3+\frac{736 \pi ^4}{45}
\]+\\
\nn&&\hat g^{12}\[
\frac{131072 L^5}{15}+\frac{327680 L^4}{3}+\left(\frac{1048576}{3}-\frac{1097728 \pi
   ^2}{9}\right) L^3\right.\\
\nn&&\;\;\;\;\;\;+L^2 \left(\frac{212992 \zeta_3}{3}+81920+24576 \pi ^2\right)\\
\nn&&\;\;\;\;\;\;+L
   \left(212992 \zeta_3-\frac{1515520}{3}+\frac{1776640 \pi ^2}{27}+\frac{39424 \pi
   ^4}{15}\right)\\
\nn&&\left.\;\;\;\;\;\;+\frac{124928 \zeta_5}{5}+\frac{106496 \zeta_3}{3}+\pi ^2
   \left(-\frac{93184 \zeta_3}{9}-\frac{107008}{27}\right)-\frac{1159424 \pi
   ^4}{675}-\frac{295936}{3}
\]+\\
\nn&&\hat g^{14}\[
\frac{1048576 L^6}{45}+\frac{6815744 L^5}{15}+\left(2752512-\frac{15007744 \pi
   ^2}{9}\right) L^4\right.\\
\nn&&\;\;\;\;\;\;+L^3 \left(\frac{3407872 \zeta_3}{9}+\frac{15073280}{3}-\frac{11141120 \pi ^2}{9}\right)\\
\nn&&\;\;\;\;\;\;+L^2 \left(2555904 \zeta_3-\frac{6914048}{3}+\frac{17096704 \pi ^2}{9}+\frac{7221248 \pi ^4}{45}\right)\\
\nn&&\;\;\;\;\;\;+L
   \left(\pi ^2 \left(\frac{61534208}{81}-\frac{5324800 \zeta_3}{9}\right)+\frac{9797632
   \zeta_3}{3}+\frac{1998848 \zeta_5}{5}-\frac{23560192}{3}-\frac{34199552 \pi
   ^4}{225}\right)\\
\nn&&\;\;\;\;\;\;+499712 \zeta_5+\pi ^2 \left(106496 \zeta_3-\frac{67858432}{243}\right)+\frac{1384448 \zeta_3^2}{9}-\frac{1384448 \zeta_3}{3}-\frac{122624 \pi ^6}{945}\\
\nn&&\left.\;\;\;\;\;\;+\frac{274300928 \pi ^4}{10125}-\frac{4759552}{15}\right]
\eeqa
where as before
$L \equiv  \log
\sqrt{8 e^{\gamma } \pi  \hat g^2}
$.
We notice that all the terms in \eq{omegapert} without $T$
are reproduced perfectly by the above expansion.

In the next section we will show how to rewrite this finite difference `boundary' problem
into a spectral problem of a Schr\"odinger equation by performing a kind of Mellin transformation.

\subsection{Equivalence to the Schr\"odinger equation}
\label{sec:schrmain}
The double scaling limit of the quark--anti-quark potential
has a long history. In \cite{Erickson:1999qv,Correa:2012nk} it was shown
that in this limit only the ladder diagrams contribute and they can
be resummed by a Bethe-Salpeter equation.
This problem can be reformulated as a problem
of finding the  ground-state energy of the Schr\"odinger  equation
\beq\la{ddF}
F''(z)+F(z)\(
\frac{4\hat g^2}{z^2+1}
-\frac{\Omega^2}{4}
\)=0\;.
\eeq
The Schr\"odinger wavefunction is linked to the solution of the Bethe-Salpeter equation.
In this section we will show that this problem is equivalent to the
second order finite difference equation arising from the QSC
accompanied by the ``quantization condition" at the
origin \eq{quantization}.

\paragraph{Relating $q$-function to the wave function.}
First we relate the $q$-function $q_1$ with the solution of \eq{ddF}
decaying at $+\infty$.
We assume that the solution decaying at $+\infty$ is normalized so that
\beq\la{normal}
F(z)\simeq e^{-\Omega z/2}\;.
\eeq
Let us show that the solution $q_1$ of \eq{bax2}
is given by the following integral Mellin-like
transformation
\beq\la{qF}
\frac{q_1(u)}{u}=2\int\limits_{i}^{+\infty}\frac{e^{-\frac{\Omega z}{2}}}{{z^2+1}}
\(\frac{z+i}{z-i}\)^{iu}
F(z)dz\;\;,\;\;{{\rm Im}\;u>0}\;.
\eeq
To see that that the equation \eq{bax2} is indeed satisfied
we consider an integral of a total derivative:
\beq
2\int\limits_{i}^{\infty}\d_z\(\[(z^2+1)F'(z)+\frac{1}{2}F(z)(-4u+\Omega+\Omega z^2)\]\frac{e^{-\frac{\Omega z}{2}}}{{z^2+1}}
\(\frac{z+i}{z-i}\)^{iu}
\)dz
\eeq
the boundary terms vanish for ${\rm Im}\;u>1$ and the integral is zero.
At the same time evaluating the derivative and
excluding the second derivative $F''(u)$
using \eq{ddF} we get
\beqa\nn
0&=&2\int\limits_{i}^{\infty}\[
(-4\hat g^2+2\Omega u-2u^2)
+u(u+i)\(\frac{z-i}{z+i}\)
+u(u-i)\(\frac{z+i}{z-i}\)
\]\frac{F(z)e^{-\frac{\Omega z}{2}}}{{z^2+1}}
\(\frac{z+i}{z-i}\)^{iu}
dz\\&=&
(-4\hat g^2+2\Omega u-2u^2)\frac{q_1(u)}{u}
+u(u+i)\frac{q_1(u+i)}{u+i}
+u(u-i)\frac{q_1(u-i)}{u-i}\;,
\eeqa
which shows that $q_1(u)$ defined by the integral \eq{qF}
satisfies \eq{bax2}.
At the same time it is easy to see by the
  saddle-point analysis that
$F(z)\simeq e^{-\Omega z/2}$
implies the following large $u$ asymptotics for $q_1$:
\beq
q_1(u)\simeq \sqrt{{\pi}/{2}}\sqrt[1/4]{8\Omega u}\;e^{-\sqrt{8\Omega u}}\;.
\eeq
Note that this map from $F(z)$ to $q_1(u)$ is valid for any (positive) value of $\Omega$.
Clearly, we have to additionally impose the decay of $F(z)$ at $z\to-\infty$ to constrain $\Omega$.
At the same time from the QSC point of view we should impose on $q_1$ the
condition \eq{quantization} at the origin.
Below we show that these two conditions are equivalent.

\paragraph{Equivalence of the two quantization conditions.}
We should relate the behavior of $q_1(u)$
near the origin with the normalizability of $F(z)$ as a solution of the
Schr\"odinger equation.
It is clear that the singularity in $q(u)/u$
around $u=0$ is due to the divergence in the integral \eq{qF} near $z=i$.
Therefore it is controlled by the behavior of $F(z)$ at $z=i$.
So our problem seems to be rather nontrivial as we have to relate the values of $F$ at large $z$
with its behavior near $z=i$. In general that would be impossible to do
without an explicit solution. However, we noticed an interesting duality of the equation which allows to do this.

The key observation is that for the normalizable $F(z)$ its Fourier image satisfies essentially the same
differential equation.
More precisely, defining $G(k)$ as
\beq\la{GF}
\frac{G(k)}{k^2+1}=\frac{\Omega^{3/2}}{8\hat g\sqrt\pi}\int\limits_{-\infty}^\infty dz F(z) e^{ik \frac{\Omega}{2} z}
\eeq
it is easy to see that $G(k)$ satisfies literally the same Schr\"odinger equation \eq{ddF}.
Furthermore, $G(k)$ also decays exponentially at both infinitites as Fourier transform of a smooth function
and is also smooth since $F$ itself decays exponentially at both infinities. This means that $F$ and $G$
should in fact coincide up to a constant factor.
To make the symmetry more manifest we can write the relation \eq{GF} between $F$ and $G$  as
\beq
F(z)=\frac{2\hat g}{\sqrt{\pi\Omega}}
\int\limits_{-\infty}^\infty dk \frac{G(k)}{k^2+1} e^{ik \frac{\Omega}{2} z}\;\;,\;\;\
G(k)=\frac{2\hat g}{\sqrt{\pi\Omega}}\int\limits_{-\infty}^\infty dz \frac{F(z)}{z^2+1} e^{ik \frac{\Omega}{2} z}\;.
\eeq
We see that in the normalization \eq{GF}\footnote{There is a possibility that $G(z)=-F(z)$, however, it is easy to see
that since $F(z)>0$ for real $z$ so must be $G$} we must have $G(z)=F(z)$,
so that we get
\beq
F(k)=\frac{2\hat g}{\sqrt{\pi\Omega}}\int\limits_{-\infty}^\infty dz \frac{F(z)}{z^2+1} e^{ik \frac{\Omega}{2} z}\;.
\eeq
This property of the solution $F(z)$ allows to bootstrap the behavior at infinity and
near the branch point $z=i$.
Let's assume that $F(z)$ has the following expansion near $z=i$:
\beqa\la{Fz}
F(z)=-\frac{i C}{2 \hat g^2}+C(z-i)\log(iz+1)+\dots
\eeqa
which is obtained by solving the equation \eq{ddF} in the vicinity of $z=i$.
As $z=i$ is the closest to the real axis singularity of $F(z)$
it controls the large $z$ behavior of $F(z)$
\beq
F(k)\simeq \frac{2\hat g}{\sqrt{\pi\Omega}}\int\limits_{-\infty}^\infty
dw \frac{-\frac{i C}{2 \hat g^2}}{w^2+1} e^{ik \frac{\Omega}{2} w}
=\frac{-i C}{\hat g}\sqrt\frac{\pi}{\Omega}   e^{-\frac{k \Omega }{2}}
\eeq
next using the normalization \eq{normal} we find
\beq
{C}= i\frac{\hat g \sqrt{\Omega}}{\sqrt{\pi }}\;,
\eeq
which fixes the expansion \eq{Fz} near $z=i$.
This allows to find the residue of $q_1(u)/u$ at the origin
by plugging \eq{Fz} into \eq{qF}:
\beq
\frac{q_1(u)}{u}\simeq \frac{i C e^{-\frac{i \Omega }{2}}}{2 \hat g^2 u}=
-\frac{1}{u}\frac{ e^{-\frac{i \Omega }{2}} \sqrt{\Omega }}{2\hat  g \sqrt{\pi }}\;.
\eeq
In Appendix \ref{sec:approof} we describe how to use a similar technique to establish the subleading
coefficient in $u$ which then gives:
\beq
\label{qqexpmain}
e^{2\pi u}q_1(u)\bar q_1(u)= -\frac{C^2}{4\hat g^4}-\frac{C^2\Omega}{8\hat g^6}u+{\cal O}(u^2)
=\frac{\Omega}{4\pi \hat g^2}
+\frac{u\Omega^2}{8\pi \hat g^4}+{\cal O}(u^2)
\eeq
showing that the condition \eq{quantization}, coming from the depth of QSC, does hold!
This finishes the proof of equivalence between the QSC and the Schr\"odinger equation in the ladders limit.

\section{Numerical solution in a wide range of the coupling}

\begin{figure}[t]{
\label{numfig}
\begin{tabular}{cc}
\includegraphics[scale=1.1]{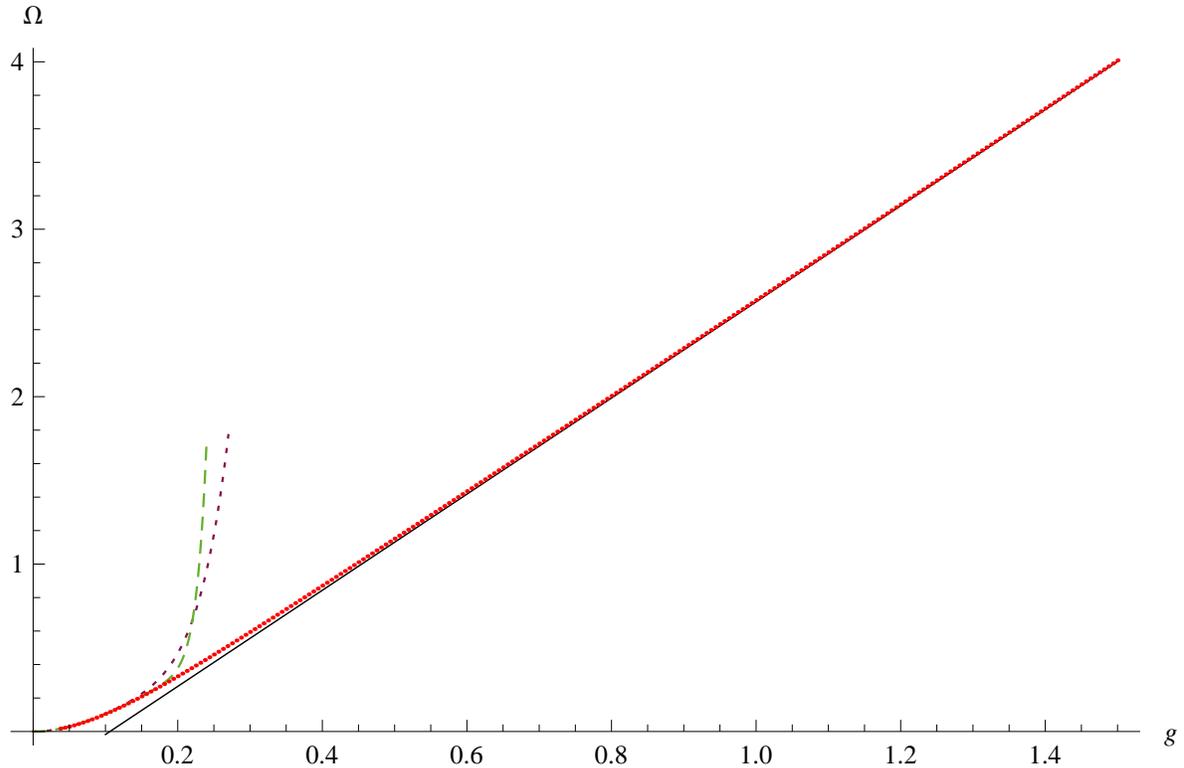}
\end{tabular}
\caption{{\textbf{Numerical results for the quark--anti-quark potential $\Omega(g)$ at $\theta=0$.}} Our numerical data points are shown in red, while the solid black line
shows the strong coupling analytic prediction \eq{strongan}. The purple curve is the 3-loop weak coupling expansion, and the dashed green curve is our 7-loop perturbative result.
}
}
\end{figure}

The QSC can be very efficiently solved numerically with essentially arbitrary precision at finite values of the coupling and all other parameters.
The general method, which is also applicable here, was developed in \cite{Gromov:2015wca}.
We have used it to generate numerical values for the quark--anti-quark potential in a wide range of the 't Hooft coupling
with $\sim 20$ digits precision. Our method works well for arbitrary real $\theta$, but we decided to focus on the case $\theta=0$. Our numerical data is listed in Appendix \ref{sec:numdata}. A plot of our results is shown on Fig. \ref{numfig}.

Let us make a comparison with the known analytical predictions. At strong coupling the classical \cite{Maldacena:1998im,Rey:1998ik} and 1-loop \cite{Forini:2010ek,Chu:2009qt} string theory results read
\beqa
\label{strongan}
\Omega\simeq\frac{\pi  (4 \pi  g+{a_1})}{4 K\left(\frac{1}{2}\right)^2}=2.8710800442 g-0.3049193809\;.
\eeqa
At the same time a fit of our numerical data gives
\beq
\Omega=2.8710800436
   g-0.3049193819+\frac{0.0100740}{g}+\frac{0.000381}{g^2}+\dots
\eeq
which quite convincingly reproduces the first two known orders.

At weak coupling
one can see on the plot that this expansion matches well our numerics. In addition, our analytic solution of the QSC at weak coupling described in Sec. \ref{sec4} provides the expansion of $\Omega$ to first $7$ loop orders presented in
\eq{omegapert} and in Appendix \ref{loop6}. Fixing a particular small value of the coupling $g=0.0625$ we compared our numerical prediction $\Omega=0.04472043670132964806$ at this point with the analytic weak coupling expansion. In Table \ref{numantab} one clearly sees that including more and more orders in the expansion improves noticeably the agreement with our numerical result. This is a nice check of our weak coupling analytic prediction.



\begin{table}[h]
\begin{center}
\begin{tabular}{|l|c|c|c|}
\hline
 & $\Omega^{\rm perturbative}$ & $\Omega^{\rm numerical}$ & $|{\rm difference}|$\\
\hline
{\rm 1-loop}  & { 0.04}908738521 & { 0.04}472043670 &  0.00436694851 \\
{\rm 2-loop}  & { 0.044}87846353 & { 0.04}472043670 &  0.00015802682 \\
{\rm 3-loop}  & { 0.0447}3327069 & { 0.04}472043670 &  0.00001283399 \\
{\rm 4-loop}  & { 0.0447}1883557 & { 0.04}472043670 &  0.00000160113 \\
{\rm 5-loop}  & { 0.044720}38490 & { 0.04}472043670 &  0.00000005179 \\
{\rm 6-loop}  & { 0.04472043}227 & { 0.04}472043670 &  0.00000000442 \\
{\rm 7-loop}  & { 0.04472043}747 & { 0.04}472043670 &  0.00000000076 \\ \hline
\end{tabular}
\end{center}
\caption{Comparison between the 7-loop weak coupling prediction and the numerical data for the quark--anti-quark potential at $g=0.0625$.}
\label{numantab}
\end{table}


\section{Conclusion}

In this paper we demonstrated that the Quantum Spectral Curve approach allows to deeply explore the quark--anti-quark potential in a variety of settings.
In particular, we generated highly precise numerical data at finite coupling interpolating extremely well between gauge theory and string theory predictions.
Thus finally we are able to access on a fully nonperturbative level this observable which historically has been a milestone in the investigations
of AdS/CFT.

The setup we study corresponds to a singular limit $\phi\to\pi$ of the cusp anomalous dimension which leads to a drastic change of Q-functions' asymptotics in the QSC.
The asymptotics we found are of a novel type even for integrable systems with twisted boundary conditions.  As this is yet another set of nontrivial asymptotics
in the QSC, it is clearly an important question how to classify all possible types of asymptotics. They should correspond to some kind of deformations
and boundary problems for local or nonlocal observables likely including the setups studied in \cite{Bajnok:2013wsa,Bajnok:2015kfz}. Consistency of asymptotics with the functional QSC equations appears to be
a highly nontrivial constraint giving hope for an exhaustive description.

Using the efficient iterative procedure of \cite{Gromov:2015vua} we computed the weak coupling expansion of the potential to the $7$th loop order.
The perturbative expansion is known to be rather nontrivial and to be captured by an effective theory arising at low energy scales. Remarkably,
we also observed the appearance of several distinct scales in the QSC which may be thought of as a counterpart
to this effective field theory description. In the future it will be also interesting to apply the QSC to study the energies of hydrogen-like bound states in $\mathcal N=4$ SYM \cite{Caron-Huot:2014gia} which are
also related to a $\phi\to\pi$ limit. Moreover, our weak coupling results may be useful to establish connections with QCD, similarly to e.g. \cite{Grozin:2015kna} (see also e.g. \cite{Brambilla:1999xf,Brambilla:2004jw,Brambilla:2014jmp} for some details of the QCD calculations of the quark-antiquark potential).

We also studied the double scaling limit when the twist $\theta$ in the scalar sector goes to $i\infty$. We showed how the Schr\"odinger equation
arising on the field theory side from resummation of ladder diagrams is encoded in the QSC, with its wavefunction rather directly linked to the Q-functions.
We believe that this approach should also apply to a   similar   double scaling limit of $\gamma$-deformed ${\mathcal N}=4$ SYM recently proposed in~\cite{Gurdogan:2015csr}, where the QSC has many common features with the one for the
cusped Wilson lines setup \cite{Gromov:2013qga,Gromov:2015dfa,Kazakov:2015efa}\footnote{The Y-system and TBA for the spectrum in the $\gamma$-deformed case were proposed earlier in \cite{Gromov:2010dy,Arutyunov:2010gu,Ahn:2011xq,deLeeuw:2012hp}}.
This limit in the $\gamma$-deformed model was   advocated in \cite{Gurdogan:2015csr} to give a novel integrable 4d theory.

We also observed a peculiar duality of the Schr\"odinger equation with respect to Fourier transform, whose meaning in the QSC itself
 beyond this special limit calls for further clarification
and might have something to do with dual conformal symmetry.
Viewing the relation between the QSC and the Schr\"odinger equation as a kind of ODE/IM correspondence \cite{Dorey:2007zx},
it would be interesting to see what kind of generalization will take place at finite twist. Another important direction is to derive the Schr\"odinger equation of \cite{Correa:2012nk} (see also \cite{Henn:2012qz}) in the ladders limit with generic $\phi$.

Finally, as the ladders limit allows for a simpler
 access to the wrapping corrections, it could also serve
as a useful ground to attempt a finite-size resummation of perturbation theory for 3-point correlators
\cite{Escobedo:2010xs,Gromov:2012vu,Vieira:2013wya,Caetano:2014gwa,Basso:2015zoa},
using Q-functions as building blocks.

\section*{Acknowledgements}

We thank
N. Drukker,
V. Kazakov,
A. Sever,
G. Sizov
and K. Zarembo for discussions. The research leading to these results has received funding from the People Programme
(Marie Curie Actions) of the European Union's Seventh Framework Programme FP7/2007-
2013/ under REA Grant Agreement No 317089 (GATIS).
 We wish to thank
STFC for support from Consolidated
grant number ST/J002798/1. The work of F.L-M. was also supported by the grant ``Exact Results in Gauge and String Theories'' from the Knut and Alice Wallenberg foundation.

\appendix
\section{Weak coupling limit of the coefficients}
\la{sec:WC}

At weak coupling one can fix the values of the several leading coefficients $A_n,B_n$ which parameterize the $\bP$-functions via \eq{fgAB}. In order to do this we used the leading order weak coupling solution of the QSC constructed in \cite{Gromov:2015dfa}. With the $\bP_a$ and $\mu_{ab}$ functions from that paper one can build $\tilde\bP_a=\mu_{ab}\chi^{bc}\bP_c$ and compare the result with our ansatz \eq{fgAB} in which $\tilde\bP_a$ is constructed by simply replacing $x\to 1/x$. For the case $\phi=\pi$ we found that
\beq
	B_0=\frac{\cos\theta-2}{\sin\theta}+A_1 g^2
\eeq
and the remaining coefficients to the leading order are all fixed as
\beqa
A_n&=&\frac{2^{n-1}\pi^n (1+(-1)^n)}{(n+1)!}+{\cal O}(g^2)\\
B_{2n}&=&\frac{(2 \pi )^{2 n-2}}{(2 n-1)!}+{\cal O}(g^2)\;\;,\;\;n>1\\
B_{2n-1}&=&-\cot\theta\frac{(2 \pi )^{2 n-2}}{(2 n-1)!}+{\cal O}(g^2)\;\;,\;\;n>1\\
B_{1}&=&2\tan\frac{\theta}{2}+{\cal O}(g^2)\\
B_{2}&=&0+{\cal O}(g^2)\ .
\eeqa

\section{Determinants entering the $5$th order equation on $\bQ_i$}
\la{sec:dets}

The 4th order difference equation \eq{bax5} on $\bQ_i$ includes several determinants built out of the $\bP$-functions, which are defined as follows:

\beqa
D_0&=&{\rm det}
\(
\bea{llll}
\bP^{1[+2]}&\bP^{2[+2]}&\bP^{3[+2]}&\bP^{4[+2]}\\
\bP^{1}&\bP^{2}&\bP^{3}&\bP^{4}\\
\bP^{1[-2]}&\bP^{2[-2]}&\bP^{3[-2]}&\bP^{4[-2]}\\
\bP^{1[-4]}&\bP^{2[-4]}&\bP^{3[-4]}&\bP^{4[-4]}
\eea
\)\;,\\
D_1&=&{\rm det}
\(
\bea{llll}
\bP^{1[+4]}&\bP^{2[+4]}&\bP^{3[+4]}&\bP^{4[+4]}\\
\bP^{1}&\bP^{2}&\bP^{3}&\bP^{4}\\
\bP^{1[-2]}&\bP^{2[-2]}&\bP^{3[-2]}&\bP^{4[-2]}\\
\bP^{1[-4]}&\bP^{2[-4]}&\bP^{3[-4]}&\bP^{4[-4]}
\eea
\)\;,\\
D_2&=&{\rm det}
\(
\bea{llll}
\bP^{1[+4]}&\bP^{2[+4]}&\bP^{3[+4]}&\bP^{4[+4]}\\
\bP^{1[+2]}&\bP^{2[+2]}&\bP^{3[+2]}&\bP^{4[+2]}\\
\bP^{1[-2]}&\bP^{2[-2]}&\bP^{3[-2]}&\bP^{4[-2]}\\
\bP^{1[-4]}&\bP^{2[-4]}&\bP^{3[-4]}&\bP^{4[-4]}
\eea
\)\;,\\
\bar D_1&=&{\rm det}
\(
\bea{llll}
\bP^{1[-4]}&\bP^{2[-4]}&\bP^{3[-4]}&\bP^{4[-4]}\\
\bP^{1}&\bP^{2}&\bP^{3}&\bP^{4}\\
\bP^{1[+2]}&\bP^{2[+2]}&\bP^{3[+2]}&\bP^{4[+2]}\\
\bP^{1[+4]}&\bP^{2[+4]}&\bP^{3[+4]}&\bP^{4[+4]}
\eea
\)\;,\\
\bar D_0&=&{\rm det}
\(
\bea{llll}
\bP^{1[-2]}&\bP^{2[-2]}&\bP^{3[-2]}&\bP^{4[-2]}\\
\bP^{1}&\bP^{2}&\bP^{3}&\bP^{4}\\
\bP^{1[+2]}&\bP^{2[+2]}&\bP^{3[+2]}&\bP^{4[+2]}\\
\bP^{1[+4]}&\bP^{2[+4]}&\bP^{3[+4]}&\bP^{4[+4]}
\eea
\)\;.
\eeqa

\section{Six and seven loop results at weak coupling}
\la{loop6}
Using the QSC we have computed the weak coupling expansion of the quark--antiquark potential at the first seven nontrivial orders. The first five orders are given in the main text in \eq{omegapert}. Here we present the rather bulky 6- and 7-loop results.

\paragraph{$6$-loop result.} The term of order $\hat g^{12}$ in $\frac{\Omega}{4\pi}$ reads
\beqa
&&\frac{131072 L^5}{15}     +      \frac{327680 L^4}{3}     +
      \frac{131072}{9} \pi ^2 L^4 T     +
   \frac{1048576 L^3}{3}     -\frac{1097728}{9} \pi ^2 L^3\\
\nn&&+      \frac{1163264}{3}
   \pi ^2 L^3 T
  +      \frac{32768}{9} \pi ^4 L^3 T^2     +      81920
   L^2     +      24576 \pi ^2 L^2     +      \frac{212992 \zeta _3 L^2}{3}     \\
\nn&&-\frac{8192}{3} \pi ^2 L^2 T     +      81920 \zeta _3 L^2 T
   -\frac{77824}{5} \pi ^4 L^2 T     +      \frac{475136}{9} \pi ^4 L^2 T^2    \\
\nn&&+
   \left(\frac{65536}{3} \pi ^2 \zeta _3 -5120 \zeta _5  \right)L^2 T^2     +      -\frac{1515520
   L}{3}     +      \frac{1776640 \pi ^2 L}{27}     +      212992 \zeta _3 L     \\
\nn&&+      \frac{39424 \pi ^4
   L}{15}     -251904 \pi ^2 L T     +      176128 \zeta _3 L T     -\frac{16384}{27} \pi ^4 L T     \\
\nn&&+      \left(10240 \zeta _5 -\frac{71680}{9} \pi ^2 \zeta _3
   \right)L T     -\frac{118784}{9} \pi ^4 L T^2     +      \left(\frac{573440}{3} \pi ^2 \zeta
   _3 -99840 \zeta _5 \right)L T^2     \\
\nn&&+      \left(3072 \zeta _3^2 +\frac{70912}{405} \pi ^6 \right)L T^2     +
\left(\frac{139264}{45} \pi ^4 \zeta _3 -31232 \pi ^2 \zeta _5  +60928 \zeta
   _7  \right) L T^3    \\
\nn&&      -\frac{295936}{3}           -\frac{107008 \pi ^2}{27}     +      \frac{106496 \zeta
   _3}{3}           -\frac{1159424 \pi ^4}{675}     +      \left(\frac{124928 \zeta _5}{5}-\frac{93184 \pi ^2 \zeta
   _3}{9}\right)     \\
\nn&&      -\frac{1190528 \pi ^2 T}{27}     -19456 \zeta _3 T     +
   \frac{3045376 \pi ^4 T}{405}     +      \left(\frac{212992}{3} \pi ^2 \zeta _3 +27648 \zeta _5
   \right)T     \\
\nn&&+      \left(1536 \zeta _3^2-\frac{14464 \pi ^6 }{405}\right)T     -\frac{50176}{3}
   \pi ^4 T^2     -      \left(\frac{172288}{3} \pi ^2 \zeta _3 +24320 \zeta _5 \right)T^2     \\
\nn&&+
   \left(18816 \zeta _3^2+\frac{17344 \pi ^6 }{135}\right) T^2     +      \left(\frac{72704}{45} \pi ^4 \zeta
   _3 +19136 \pi ^2 \zeta _5 -38976 \zeta _7 \right) T^2     \\
\nn&&+      \left(\frac{228352}{45} \pi ^4 \zeta _3
   -\frac{107264}{3} \pi ^2 \zeta _5 +43904 \zeta _7 \right)T^3\\
\nn&&     +      \left(2496  \zeta
   _{6,2}+\frac{20224}{3} \pi ^2 \zeta _3^2 -31232 \zeta _3 \zeta _5 +\frac{55304 \pi ^8
   }{42525}\right)T^3     \\
\nn&&+      \left(-\frac{2560}{3} \pi ^4 \zeta _5 +21504 \pi ^2 \zeta _7 -102816
   \zeta _9 \right)T^4
\eeqa
At this order an irreducible multiple zeta value appears for the first time, given by $\zeta_{6,2}\simeq 0.017819740416836$.

\paragraph{$7$-loop result.}
The term of order $\hat g^{14}$ in $\frac{\Omega}{4\pi}$ is given by
{\footnotesize
\beqa
&&\frac{1048576 L^6}{45}+\frac{524288}{9} L^5 \pi ^2 T+\frac{6815744 L^5}{15}+\frac{262144}{9} L^4 \pi ^4 T^2-65536 L^4 T \zeta_3+\frac{40632320}{9} L^4 \pi ^2
   T\\
\nn&&-\frac{15007744}{9} L^4 \pi ^2+2752512 L^4+\frac{131072}{81} L^3 \pi ^6 T^3+65536 L^3 \pi ^2 T^2 \zeta_3+\frac{655360}{3} L^3 T^2 \zeta_5\\
\nn&&+\frac{12255232}{9} L^3 \pi ^4
   T^2-\frac{64159744}{135} L^3 \pi ^4 T-65536 L^3 T \zeta_3+\frac{13303808}{3} L^3 \pi ^2 T+\frac{3407872 L^3 \zeta_3}{9}\\
\nn&&-\frac{11141120}{9} L^3 \pi ^2+\frac{15073280
   L^3}{3}+\frac{2080768}{45} L^2 \pi ^4 T^3 \zeta_3-\frac{499712}{3} L^2 \pi ^2 T^3 \zeta_5-129024 L^2 T^3 \zeta_7\\
\nn&&+32768 L^2 \pi ^6 T^3-\frac{2828288}{405} L^2 \pi ^6
   T^2-36864 L^2 T^2 \zeta_3^2+\frac{11444224}{3} L^2 \pi ^2 T^2 \zeta_3+20480 L^2 T^2 \zeta_5\\
\nn&&+\frac{2351104}{3} L^2 \pi ^4 T^2-\frac{7610368}{9} L^2 \pi ^2 T \zeta_3-40960
   L^2 T \zeta_5-\frac{27344896}{45} L^2 \pi ^4 T+1671168 L^2 T \zeta_3\\
\nn&&-3817472 L^2 \pi ^2 T+\frac{7221248 L^2 \pi ^4}{45}+2555904 L^2 \zeta_3+\frac{17096704 L^2 \pi
   ^2}{9}-\frac{6914048 L^2}{3}+\frac{8192}{9} L \pi ^6 T^4 \zeta_3\\
\nn&&-\frac{133120}{3} L \pi ^4 T^4 \zeta_5+369152 L \pi ^2 T^4 \zeta_7-628992 L T^4 \zeta_9+\frac{1176832 L
   \pi ^8 T^3}{42525}+\frac{210944}{3} L \pi ^2 T^3 \zeta_3^2\\
\nn&&-71680 L T^3 \zeta_3 \zeta_5+30720 L T^3 \zeta_{6,2}+\frac{7872512}{15} L \pi ^4 T^3 \zeta_3-1899520 L \pi ^2
   T^3 \zeta_5+867328 L T^3 \zeta_7\\
\nn&&+\frac{212992}{27} L \pi ^6 T^3-\frac{1150976}{15} L \pi ^4 T^2 \zeta_3+665600 L \pi ^2 T^2 \zeta_5-268800 L T^2 \zeta_7+\frac{2378752}{405} L \pi ^6 T^2\\
\nn&&+43008 L T^2 \zeta_3^2+\frac{757760}{3} L \pi ^2 T^2 \zeta_3-1587200 L T^2 \zeta_5-\frac{14838784}{9} L \pi ^4 T^2-\frac{2152448 L \pi
   ^6 T}{2835}\\
\nn&&-163840 L T \zeta_3^2+\frac{24051712}{9} L \pi ^2 T \zeta_3+364544 L T \zeta_5+\frac{390412288}{405} L \pi ^4 T+2457600 L T \zeta_3\\
\nn&&-\frac{39706624}{9} L \pi ^2
   T-\frac{5324800}{9} L \pi ^2 \zeta_3+\frac{1998848 L \zeta_5}{5}-\frac{34199552 L \pi ^4}{225}+\frac{9797632 L \zeta_3}{3}\\
\nn&&+\frac{61534208 L \pi ^2}{81}-\frac{23560192
   L}{3}-\frac{11264}{105} \pi ^6 T^5 \zeta_5+\frac{73216}{5} \pi ^4 T^5 \zeta_7-285120 \pi ^2 T^5 \zeta_9\\
\nn&&+1271952 T^5 \zeta_{11}-\frac{10544 \pi ^{10}
   T^4}{93555}+\frac{91136}{9} \pi ^4 T^4 \zeta_3^2-\frac{520832}{3} \pi ^2 T^4 \zeta_3 \zeta_5+179424 T^4 \zeta_5^2\\
\nn&&+361088 T^4 \zeta_3 \zeta_7+\frac{16768}{3} \pi ^2 T^4
   \zeta_{6,2}-26432 T^4 \zeta_{8,2}+\frac{65536}{45} \pi ^6 T^4 \zeta_3-63488 \pi ^4 T^4 \zeta_5\\
\nn&&+401408 \pi ^2 T^4 \zeta_7-508032 T^4 \zeta_9+\frac{5137792 \pi ^6 T^3
   \zeta_3}{2835}-768 T^3 \zeta_3^3+30976 \pi ^4 T^3 \zeta_5\\
\nn&&-\frac{941632}{3} \pi ^2 T^3 \zeta_7+\frac{2211904 T^3 \zeta_9}{3}
-\frac{142816 \pi ^8
   T^3}{14175}+\frac{1183232}{3} \pi ^2 T^3 \zeta_3^2-337664 T^3 \zeta_3 \zeta_5\\
\nn&&+17664 T^3 \zeta_{6,2}-\frac{256000}{3} \pi ^4 T^3 \zeta_3+\frac{1762304}{3} \pi ^2 T^3 \zeta_5+367360 T^3 \zeta_7-\frac{446464}{45} \pi ^6 T^3\\
\nn&&+\frac{2348512 \pi ^8 T^2}{42525}-\frac{175360}{3} \pi ^2 T^2 \zeta_3^2+\frac{76288}{3} T^2 \zeta_3 \zeta_5+26880 T^2
   \zeta_{6,2}+\frac{6986752}{45} \pi ^4 T^2 \zeta_3\\
\nn&&+\frac{295424}{9} \pi ^2 T^2 \zeta_5-611520 T^2 \zeta_7-\frac{1111552}{405} \pi ^6 T^2+225792 T^2 \zeta_3^2-\frac{2234624}{3} \pi ^2 T^2 \zeta_3\\
\nn&&-261120 T^2 \zeta_5+\frac{3700736 \pi ^4 T^2}{27}-\frac{2342656}{135} \pi ^4 T \zeta_3+131584 \pi ^2 T \zeta_5+33152 T \zeta_7\\
\nn&&+\frac{3462656 \pi ^6 T}{2835}+165888 T \zeta_3^2+\frac{3972608}{27} \pi ^2 T \zeta_3+387072 T \zeta_5-\frac{222660352 \pi ^4 T}{1215}\\
\nn&&-544768 T \zeta_3+\frac{16618240
   \pi ^2 T}{81}-\frac{122624 \pi ^6}{945}+\frac{1384448 \zeta_3^2}{9}+106496 \pi ^2 \zeta_3+499712 \zeta_5\\
\nn&&+\frac{274300928 \pi ^4}{10125}-\frac{1384448 \zeta_3}{3}-\frac{67858432 \pi ^2}{243}-\frac{4759552}{15}
\eeqa
}
where we have a new multiple zeta value $\zeta_{8,2}\simeq 0.0041224696783998322240$.

\section{Complex conjugation of the $\bQ_i$ functions}
\label{sec:compcon}
Another set of useful relations concerns the expected symmetry of the QSC system under complex conjugation.
Let's assume that under the complex conjugation the equation \eq{bax5}
remain invariant. In general this is true if
\beq
\bar \bP_{a}=\lambda_{a}^{\;\;b}\bP_b\;\;,\;\;
\bar \bP^{a}=\lambda^{a}_{\;\;b}\bP^b
\eeq
for some constant coefficients $\lambda_{a}^{\;\;b}$,
such that $\lambda_{a}^{\;\;b}\lambda^{a}_{\;\;c}=-\delta_c^b$
(in our case $\lambda_{a}^{\;\;b}=-i\delta_a^b$).
If this holds the complex conjugate $\bar {\bf Q}_i$ should give an
alternative complete set of solutions of the finite difference equation \eq{bax5},
which should be related to the initial set as a linear combinations
with some $i$-periodic coefficients
\beq\la{Qomega}
\bar {\bf Q}_i=\Omega_{i}^{\;\;j}{\bf Q}_j\;\;,\;\;\Omega_{i}^{\;\;j}(u+i)=\Omega_{i}^{\;\;j}(u)\;.
\eeq
Those coefficients can be written in terms of ${ Q}_{a|i}$ as
\beq\la{OmegaQQ}
\Omega_{i}^{\;\;j}=-\bar {Q}_{a|i}(u-\tfrac{i}{2})\lambda^{a}_{\;\;b}Q^{b|j}(u-\tfrac{i}{2})
\eeq
where $Q^{b|j}=-((Q_{b|j})^{-1})^{\rm T}$. We can easily check this is indeed
true. We show that \eq{Qomega} holds:
\beq
\Omega_{i}^{\;\;j}{\bf Q}_j=
-\bar {Q}_{a|i}(u-\tfrac{i}{2})\lambda^{a}_{\;\;b}Q^{b|j}(u-\tfrac{i}{2})
{\bf Q}_j=
-\bar {Q}_{a|i}(u-\tfrac{i}{2})\lambda^{a}_{\;\;b}{\bf P}^b
=
-\overline{ {Q}_{a|i}(u+\tfrac{i}{2}) {\bf P}^a}
=
\bar {\bf Q}_i
\eeq
and also that the r.h.s. \eq{OmegaQQ} is  periodic:
\beq
\bar {Q}_{a|i}^{+}\lambda^{a}_{\;\;b}Q^{b|j+}=
\overline{
(Q_{a|i}^+-\bP_a {\bf Q}_i)
}\lambda^{a}_{\;\;b}(Q^{b|j-}+\bP^b\bQ^j)
=
\bar
Q_{a|i}^-
\lambda^{a}_{\;\;b}Q^{b|j-}\;
\eeq
(we denoted $f^\pm=f(u\pm i/2)$).
Finally we can find discontinuity of $\Omega$
using this identity
\beq
\tilde\Omega_{i}^{\;\;j}-
\Omega_{i}^{\;\;j}=
-\bar {Q}_{a|i}^-\lambda^{a}_{\;\;b}\(\tilde Q^{b|j-}
-Q^{b|j-}\)=
\bar {Q}_{a|i}^-\lambda^{a}_{\;\;b}
\(\tilde \bP^b\tilde \bQ^j-
\bP^b \bQ^j\)=-\bar{\tilde \bQ}_i\tilde\bQ^j
+
\bar{\bQ}_i\bQ^j\;.
\eeq
We notice one more relation which we will use below. Consider $\Omega_1^{\;4}$. Its discontinuity is
due to \eq{qtil}
\beq\la{dom14}
\tilde\Omega_1^{\;4}(u)-\Omega_1^{\;4}(u)=
u\bar{\bq}_1(u)\bq_1(u)
-u\bar{\bq}_1(-u)\bq_1(-u)
\eeq
from where we see that $\Omega_1^{\;4}$ should be an even function.

\section{Expansion of $q_1(u)$ at the origin}
\label{sec:approof}
As discussed in the end of Sec \ref{sec:schrmain}, to demonstrate that the Schr\"odinger equation is encoded in the QSC we need to compute the expansion of $q_1(u)$ at the origin up to the term linear in $u$. Let us show how this can be done.

On the one hand, from the 2nd order difference equation \eq{bax2} on $q_1$ we find that $q_1(0)$ and $q_1'(0)$ are related to its expansion at $u=-i$:
\beq
\label{q1expbax}
	q_1(u)=\frac{4\hat g^2}{(u+i)^2}q_1(0)+\frac{4\hat g^2q_1'(0)-2q_1(0)\Omega}{u+i}
	+{\mathcal O}((u-i)^0)\ .
\eeq
On the other hand, we can compute the expansion around $u=-i$ using the expression \eq{qF} for $q_1$ in terms of $F(z)$. In that expression the singularity of $q_1$ at $u=-i$ arises because the integrand is singular when $z=i$. In the vicinity of this point $F(z)$ is a linear combination of two solutions of the the Schr\"odinger equation, one of which is smooth at $z=i$ and the other one also includes terms of the type $(z-i)^n\log(iz+1)$ with $n\geq 1$. Solving the equation close to this point we find
\beq
	F(z)=-\frac{iC}{2\hat g^2}+C(z-i)\log(iz+1)+iC_2(z-i)+\dots\ ,
\eeq
where the real\footnote{One can show that $C_2$ is real using the fact that $F(z)$ is a real and even function. } constant $C_2$ comes from the smooth solution and dots stand for more regular terms. Let us also note that the expression \eq{qF} is not applicable directly for ${\rm Im}\; u<0$ as the integrand is too singular near $z=i$. However, as we need only the coefficients of the double and the single pole at $u=-i$ in $q_1(u)$, we can modify \eq{qF} in a way which ensures convergence of the integral without changing these two coefficients:
\beq
\label{regint}
	2\int\limits_i^\infty dz \frac{e^{-i\Omega/2}}{z^2+1}\(\frac{z+i}{z-i}\)^{iu}
	\[\frac{iC}{2\hat g^2}+e^{-i\frac{z-\Omega}{2}}
	\(-\frac{iC}{2\hat g^2}+C(z-i)\log(iz+1)+iC_2(z-i)+\dots\)\]\ .
\eeq
We subtracted a part proportional to the integral
\beq
\label{refint}
	2\int\limits_i^\infty dz \frac{1}{z^2+1}\(\frac{z+i}{z-i}\)^{iu}=\frac{1}{u}\ ,
\eeq
which does not affect the two coefficients we are after. From \eq{regint} we now find
\beq
q_1(u)=
-\frac{2 i C e^{-\frac{i \Omega }{2}}}{(u+i)^2}
	-\frac{e^{-\frac{i \Omega }{2}} \left(-\frac{i C \Omega }{2 \hat g^2}-2 i \pi  C-2 C-2 C
   \log 2-2 i C_2\right)}{u+i}+{\mathcal O}((u-i)^0)\ .
\eeq
Comparing this with \eq{q1expbax} we get
\beq
	q_1(0)=-\frac{i C e^{-\frac{i \Omega }{2}}}{2 \hat g^2}\ ,
\eeq
\beq
	q_1'(0)=\frac{e^{-\frac{i \Omega }{2}} \left(4 \hat{g}^2 \left(i C_2+C (1+i \pi +\log 2)\right)-i
   C \Omega \right)}{8 \hat{g}^4}\ .
\eeq
This finally allows to construct the combination $e^{2\pi u}q_1(u)\bar q_1(u)$ which we need. We observe that $C_2$ cancels out and we find
\beq
	e^{2\pi u}q_1(u)\bar q_1(u)= -\frac{C^2}{4\hat g^4}-\frac{C^2\Omega}{8\hat g^6}u+{\cal O}(u^2)\ ,
\eeq
which is the key result used in \eq{qqexpmain} in the main text.

\section{Numerical data}
\label{sec:numdata}

Here we present a part of our numnerical data for the quark--antiquark potential $\Omega$ at finite coupling $g$ with zero twist $\theta=0$.
While the accuracy might vary slightly, we expect
all digits to be correct (with uncertanity in the last digit).
\beq
\nn
\begin{array}{|l|l||l|l|}
\hline
g&\Omega(g)&g&\Omega(g)
\\ \hline
 0 & 0 & 0.05 &                          0.02937069654776 \\
 0.075 & 0.06265474565224 & 0.1 & 0.10511713720337 \\
 0.125 & 0.15465836443567 & 0.15 & 0.20955607216466 \\
 0.175 & 0.26845318866584 & 0.2 & 0.330312294925133 \\
 0.225 & 0.39435828555165 & 0.25 & 0.4600215248401101992 \\
 0.275 & 0.5268878004652301086 & 0.3 & 0.5946574683022822222 \\
 0.325 & 0.6631138101939375140 & 0.35 & 0.7320994342456940408 \\
 0.375 & 0.8014991401020814198 & 0.4 & 0.8712277052055592640 \\
 0.425 & 0.9412212786455914103 & 0.45 & 1.0114313519742950991 \\
 0.475 & 1.0818205391585539063 & 0.5 & 1.1523596132795935855 \\
 0.525 & 1.2230254118796313025 & 0.55 & 1.2937993424631526624 \\
 0.575 & 1.3646663040854278314 & 0.6 & 1.4356138993330072537 \\
 0.625 & 1.5066318508313724359 & 0.65 & 1.5777115633938239593 \\
 0.675 & 1.6488457911511407735 & 0.7 & 1.7200283813289496328 \\
 0.725 & 1.7912540747152853534 & 0.75 & 1.8625183485921063555 \\
 0.775 & 1.9338172918626574100 & 0.8 & 2.0051475048695944173 \\
 0.825 & 2.0765060183502537912 & 0.85 & 2.1478902273706390145 \\
 0.875 & 2.2192978370894877695 & 0.9 & 2.2907268179434612004 \\
 0.925 & 2.3621753683925865485 & 0.95 & 2.4336418837756454947 \\
 0.975 & 2.5051249301358780128 & 1. & 2.5766232221146891288 \\
 1.025 & 2.6481356041939115734 & 1.05 & 2.7196610347092241265 \\
 1.075 & 2.7911985721684911266 & 1.1 & 2.8627473634963918184 \\
 1.125 & 2.9343066338961908685 & 1.15 & 3.0058756780749462488 \\
 1.175 & 3.0774538526229463147 & 1.2 & 3.1490405693740687669 \\
 1.225 & 3.2206352896028680533 & 1.25 & 3.2922375189379219065 \\
 1.275 & 3.3638468028903889215 & 1.3 & 3.4354627229126975867 \\
 1.325 & 3.5070848929154719081 & 1.35 & 3.5787129561817286982 \\
 1.375 & 3.6503465826264772085 & 1.4 & 3.7219854663574488362 \\
 1.425 & 3.793629323499052523 & 1.45 & 3.865277890247007235 \\
 1.475 & 3.936930921125622182 & 1.5 & 4.008588187423520918 \\
\hline
\end{array}
\eeq

\end{document}